# Noise floor reduction of an Er:fiber laser-based photonic microwave generator

**Haifeng Jiang, Jennifer Taylor, (*IEEE Student Member*), Frank Quinlan, (*IEEE Member*), Tara Fortier and Scott A. Diddams, (*IEEE Member*)**

*National Institute of Standards and Technology, Boulder, CO 80305 USA*

*Corresponding author: haifeng.jiang@nist.gov, scott.diddams@nist.gov*

**Abstract:** The generation of microwaves from optical signals suffers from thermal and shot noise inherent in the photodetection process. This problem is more acute at lower pulse repetition rates where photodiode saturation limits the achievable signal-to-noise ratio. In this paper, we demonstrate a 10-15 dB reduction in the 10 GHz phase noise floor by multiplication of the pulse repetition rate. Starting with a 250 MHz fundamentally mode-locked erbium(Er):fiber laser, we compare two different approaches to repetition rate multiplication: Fabry-Perot cavity filtering and a cascaded, unbalanced Mach-Zehnder fiber-based interferometer. These techniques reduce the phase noise floor on the 10 GHz photodetected harmonic to -158 dBc/Hz and -162 dBc/Hz, respectively, for Fourier frequencies higher than 100 kHz.

## 1. Introduction

Low phase noise microwave generation plays an important role in many applications such as radar systems [1], communication systems, deep space navigation [2], timing distribution and synchronization [3], arbitrary waveform generation and novel imaging [4], precision spectroscopy and frequency metrology [5, 6]. Photonically-generated microwaves provide unprecedented close-to-carrier phase noise and short term stability [5, 7, 8] by taking advantage of the performance of ultra-stable laser cavities [9-11] and optical frequency dividers [12, 13]. The basic structure of this type of photonic microwave generator includes: (1) a continuous-wave (CW) laser that is frequency-stabilized to an ultra-stable optical cavity, (2) a self-referenced [14] mode-locked laser that is phase-locked to the CW laser, and (3) a photodiode detecting the mode-locked laser repetition rate [7]. The mode-locked laser functions as an optical-to-microwave frequency divider, and the desired signal at the repetition frequency (or its harmonics) is selected with a microwave bandpass filter at the photodiode output. In this work, we focus on microwave generation at a harmonic of the laser repetition rate near 10 GHz, which is an important frequency for radar, communications and frequency metrology.

Self-referenced Er:fiber mode-locked lasers are commercially available, relatively compact, and robust, making them convenient to be used in such an approach to generating ultra-low phase noise microwave signals. However, when compared with other self-referenced lasers, such as a gigahertz Titanium-sapphire (Ti:S) [15], the Er:fiber laser has a lower repetition rate (a few hundred megahertz) and correspondingly higher pulse energy that result in more rapid saturation of the microwave signal from the photodiode detecting the repetition rate. As previously described [16], once the photodiode is saturated, it is not possible to further reduce the shot or thermal noise floor of the generated microwave signals by increasing the optical power. The details of the photodiode saturation [17, 18] depend on the exact repetition rate, pulse length and photodiode design. But for common parameters and devices, saturation typically begins with only ~1 mW average power for a laser with 250 MHz repetition rate. This limits the power generated in the 10 GHz harmonic to approximately -20 dBm, which currently results in a shot- or thermal-noise limited phase noise floor of about -145 dBc/Hz on the 10 GHz photodetected harmonic [19]. A mode-locked laser with larger fundamental repetition rate (e.g. 1-10 GHz) would have a definite advantage in this regard [20, 21]; however, high repetition rate, low-noise, fiber-based sources are challenging to construct, particularly when one considers the octave-spanning spectrum required for self-referencing. This apparent drawback of fiber-based systems can be overcome by multiplication of the pulse repetition rate. One approach uses a mode-filtering Fabry-Perot (FP) cavity [16], while an alternative approach uses a fiber-based, cascaded Mach-Zehnder interferometer (MZI) [22]. In this paper, we compare these two techniques, with a focus on the reduction of the microwave signal noise floor. Predicted shot and thermal noise levels for our system are also presented in order to compare the measured phase noise to these noise limits.

## 2. Thermal and Shot Noise Limitations in Photodetection

For higher Fourier frequencies (>100 kHz), the phase noise on the photodetected output of a mode-locked laser is mainly attributed to the thermal (Johnson) noise, shot noise, and laser amplitude noise converted to phase noise in the photodiode. The following is a simplified estimation of the phase noise caused by two of these sources. The analysis below is suitable for the common photodetection circuit shown in Figure 1, which consists of a reverse-biased high-speed PIN photodiode with a terminating 50 $\Omega$ resistor, that is coupled to an external 50 $\Omega$ matching circuit or component.

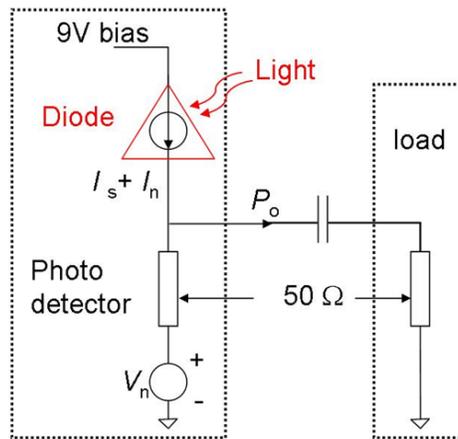

**Figure 1**. Photodetection circuit and noise model, where $I_s$ is the average DC photocurrent, $I_n$ is the shot noise current source, $P_o$ is the output AC power of the photodetector at the desired frequency, and $V_n$ is the thermal noise voltage source intrinsic to the 50 $\Omega$ resistor.

Thermal or Johnson noise arises from the thermal agitation of the charge carriers across the resistive element [23, 24]. The single-sided, double-sideband power spectral density (PSD), or voltage variance per hertz of bandwidth is given by:

$$\overline{V_n^2} = 4k_B TR \quad \text{V}^2/\text{Hz} \quad (1)$$

where $T$ is temperature in Kelvins, $k_B$ is the Boltzmann constant, and $R$ (50 $\Omega$) is the resistance. Because both source and load have 50 $\Omega$ impedance, only half of the AC source voltage drops across either of these resistors. Thus, the resulting noise power measured in the external circuit is reduced by a factor of four. The noise power is independent of the signal, so that it contributes equally to both the phase and amplitude of the 10 GHz signal. This accounts for a reduction by

another factor of two, such that the single-sideband phase noise induced by the thermal noise will be [25]

$$L(f) = 10\log\frac{k_B T}{2P_o} \approx -177 - 10\log[P_o(mW)], \quad \text{dBc/Hz} \quad (2)$$

where $P_o$ is the power across 50 $\Omega$, as measured with a spectrum analyzer at the output of the photodiode at the desired harmonics of the repetition rate.

Shot noise is a fundamental noise process in the photodetection of a light source. It is characterized by a single-sided, double-sideband PSD of current variance per hertz of bandwidth that is given by:

$$\overline{I_n^2} = 2eI_s, \quad \text{A}^2/\text{Hz} \quad (3)$$

where $e$ is the electron charge, and $I_s$ is the average direct current (DC) measured at the PD output with a voltage-meter. Here, we again assume that the noise power equally acts on both phase and amplitude of the microwave signal. Similar to the thermal noise effect, the noise power is also reduced by a factor of four due to the inner resistance of the photodiode. Therefore, the shot noise results in a single-sideband phase noise of:

$$L(f) = 10\log\frac{eI_s R}{4P_o} \approx -177 + 10\log\left[\frac{I_s(mA)}{P_o(mW)}\right]. \quad \text{dBc/Hz} \quad (4)$$

Combining Eqns (2) and (4) yields the following expression for the phase noise arising from the thermal and shot noise processes:

$$L(f) = 10\log\frac{eI_s R + 2k_B T}{4P_o}. \quad \text{dBc/Hz} \quad (5)$$

It is worth noting that below saturation, $P_o$ is proportional to $I_s^2$, such that the thermal noise floor improves at a rate of 20 dB per decade increase in average photocurrent. On the other hand, at photocurrents greater than approximately 1 mA, the shot-noise limited floor begins to dominate and improves at the lesser rate of 10 dB per decade.

Beyond these fundamental sources of noise, the conversion of laser amplitude noise (AM) to phase noise (PM) in the process of photodetection cannot be neglected [26]. The AM-to-PM conversion coefficient, $\alpha$, is strongly dependent on the specific photodiode properties [27] and the energy per laser pulse [28], and at 10 GHz it can vary from $\alpha = 0$ to $\alpha > 1$ rad/($\Delta P/P$), where $\Delta P/P$ is the fractional fluctuation in optical power [27]. As discussed further below, in many conditions, this source of phase noise can be much greater than that caused by thermal and shot noise.

## 3. Experimental Setup and Measurement System

As noted above, the use of a higher repetition rate source is a route to improving the phase noise floor of a photodetected pulse train. In this section we present two approaches to repetition rate multiplication. Figure 2 shows the scheme of a FP mode-filtering cavity. The cavity is composed of two concave mirrors (radius of curvature of 50 cm) with a reflectivity of 97 %, corresponding to a finesse of about 100. The length of the cavity is about 3 cm, corresponding to a 5 GHz free spectral range (FSR). In order to obtain a large transmission signal, we optimize the input beam profile with lenses to match the lowest order spatial cavity mode (TEM00). One mirror of the cavity is glued to a ring-shaped piezoelectric transducer (PZT), which enables cavity length control. A pick-off mirror with a reflectivity of a few percent is used to direct light reflected from the cavity onto a split detector for locking the cavity length to a harmonic of the laser repetition rate by using the tilt-lock method [29]. The resulting output of the cavity filter is a thinned frequency comb with only one out of every 20 input modes transmitted. In the time domain, this corresponds to a pulse circulating in the cavity multiple times to produce a 5 GHz repetition rate signal.

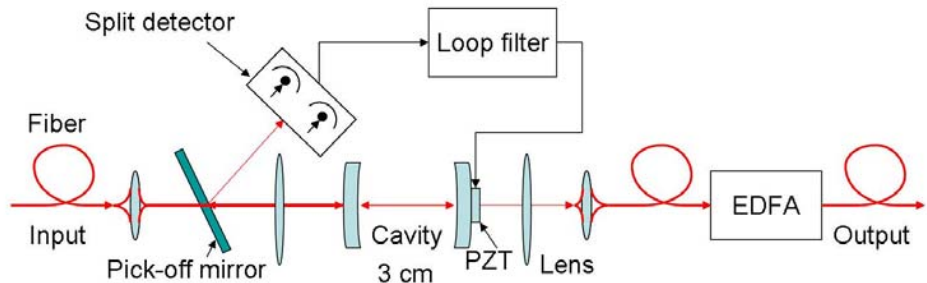

**Figure 2**. Scheme of a mode-filtering cavity, where PZT is a ring-shaped piezoelectric transducer, EDFA is erbium-doped fiber amplifier.

The main drawback of this approach is that it is inefficient. For a 250 MHz input, the 5 GHz cavity transmits only 5 % of the incident power [16]. Additional loss results from a tradeoff between high transmission of the TEM00 mode and good error signal sensitivity. The error signal of the tilt lock is obtained from interference between the TEM00 mode and a higher-order mode of the cavity. Thus, a larger slope of error signal (and higher signal-to-noise ratio of the control loop) corresponds to more power coupled into the higher-order mode. Consequently, to avoid degradation in system performance from limited signal-to-noise in the control loop, the coupling to the TEM00 mode is decreased to provide improvement of error-signal sensitivity. As a result of these effects, only 1.2 mW of optical power is coupled into the fiber after the cavity, while the incident power is 90 mW. A commercial erbium-doped fiber amplifier (EDFA) is used to subsequently increase the optical signal.

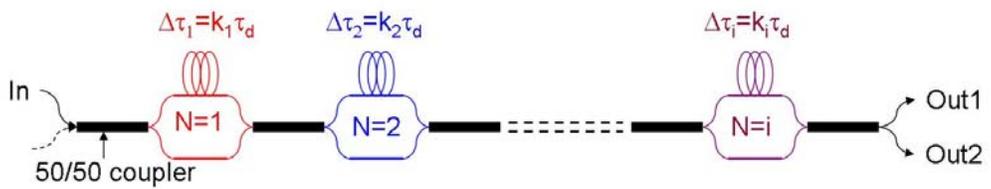

**Figure 3**. Scheme of the fiber-based cascaded Mach-Zehnder interferometer (MZI), where $k_1$, $k_2$ and $k_i$ are integers, $\tau_d$ (=100 ps) is the period of the desired frequency signal. No pulse overlapping happens at the output.

Figure 3 shows the alternative approach of repetition rate multiplication based on a series of cascaded MZI, which act to split the input and then delay and temporally interleave the pulse trains from the two interferometer arms [22, 30, 31]. Here we employ standard four-port 50/50 fiber couplers with the two outputs of each stage serving as the inputs to the next stage. As a result, even with inexpensive off-the-shelf couplers the transmission efficiency from a single output after a few stages can easily reach > 40 %, which is much greater than that of the mode-filtering cavity approach. (Note that the second output could drive a second photodiode and the signals could be summed electronically to potentially recover the benefit of the full optical power). Although we call it an interferometer, we in fact avoid direct temporal overlap of the pulses from the two arms of each stage of the MZI. This reduces significant noise related to optical phase variation. Each MZI stage increases the number of pulses in the output by a factor of two, but does not necessarily have to double the repetition rate. By selecting a set of proper values of $k_1$, $k_2$ ...and $k_i$, we can satisfy the condition of no pulses overlapping and all pulses contributing in phase to the desired 10 GHz signal. Trimming and splicing of the fiber delay in the interferometer arms is accomplished with the aid of a high-speed oscilloscope, and as discussed below, satisfactory tolerances below 5 ps (~1 mm of fiber) are readily achieved.

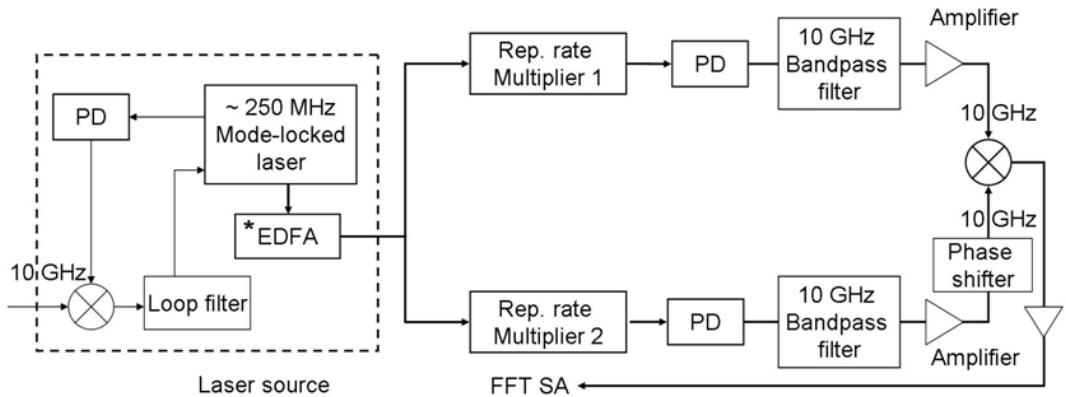

**Figure 4**. Scheme of the phase noise measurement setup, where PD is photodiode; the EDFA is used only for measuring the residual noise with the FP cavity multiplier; FFT SA is fast Fourier-transform spectrum analyzer.

To determine the impact of these repetition rate multiplication approaches on the phase noise floor of the photodetected optical signals, we compare the signals from two independent repetition rate multipliers using the residual phase noise measurement system of Figure 3. We use a commercial 250 MHz Er:fiber mode-locked laser (1550 nm center wavelength) as the source. Its repetition rate is locked to a 10 GHz reference frequency as shown in Figure 3. This laser has two 35 mW outputs, which are directly used to measure the phase noise of two cascaded MZI. For the cavity case, one of the laser outputs is amplified to ~ 180 mW, which is then split evenly and input to the two FP cavities. The outputs of the multipliers are detected with high-power 22 GHz InGaAs photodiodes having responsivity at 1550 nm of ~0.5 A/W [32]. We note that the laser bandwidth supports ~100 femtosecond pulses, but un-compensated dispersion in the laser, fiber and other components results in chirped pulses that are at maximum a few picoseconds in duration before photodetection. The photodiodes generate a comb of microwave frequencies at harmonics of the repetition rate of the optical pulse train. The desired frequency of 10 GHz is selected by use of narrow bandpass cavity-filters, and microwave amplifiers are used as needed to increase the signals to about 0 and 7 dBm to drive the RF and LO ports of the mixer, respectively. The relative phases of the two inputs to the mixer are adjusted to be in quadrature, for maximum sensitivity to phase fluctuations. A low-noise baseband amplifier with 20 dB gain is used after the mixer to enhance the measurement sensitivity, and the voltage fluctuations at its output are recorded and analyzed with a Fourier-transform spectrum analyzer. By use of the measured voltage-to-phase conversion factor of the mixer, the spectral density of phase noise is obtained.

## 4. Experimental results

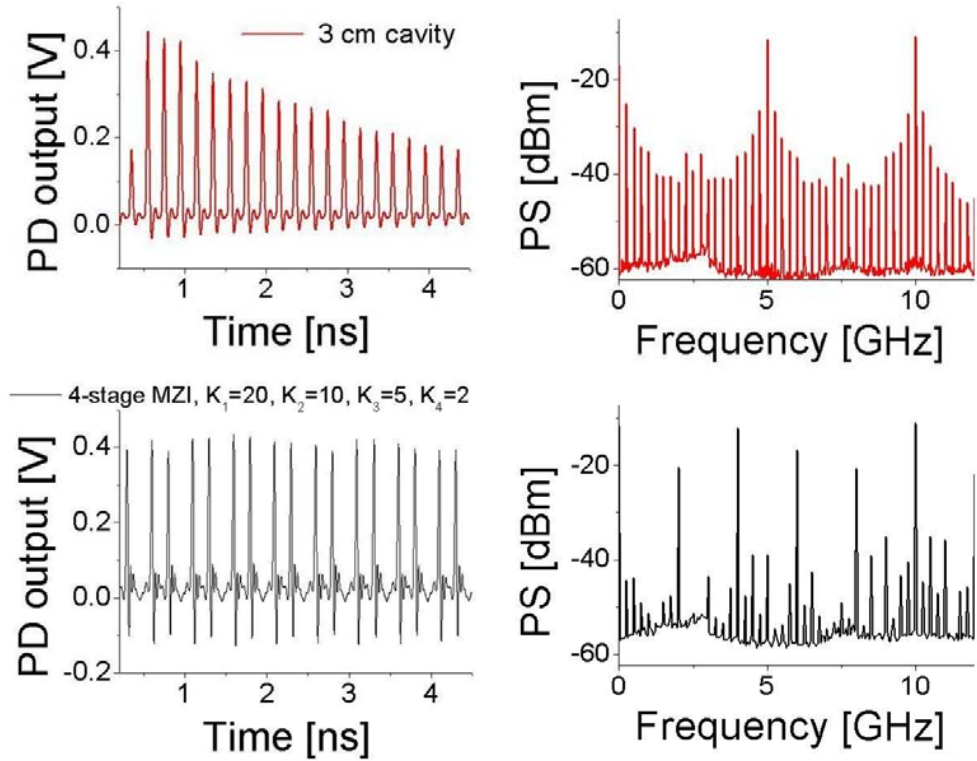

**Figure 5**. Photodiode (PD) output in time and frequency domain with (upper) 3 cm FP filter cavity, and (lower) 4-stage MZI with parameters given in the legend. The optical power in all cases is 3.5 mW. PS denotes power spectrum.

Figure 5 shows the signal of PD output in the time and frequency domains with the 3 cm FP cavity and the 4-stage MZI with $k_1$ = 20, $k_2$ = 10, $k_3$ = 5 and $k_4$ = 2. The time domain signals are measured with a 20 GHz oscilloscope, while the spectra are obtained with a spectrum analyzer. The incident power is about 3.5 mW in all cases. In the case of the FP cavity, the microwave spectrum shows suppression of unwanted modes, consistent with the finesse of the cavity. In contrast, the spectrum of the output from the 4-stage MZI is much more irregular, with the details dependent of the exact delay values in the various stages. Despite these marked differences, the 10 GHz signal size is approximately -11 dBm in both cases. This level is well in agreement with the fact that the alternating current (AC) power at a harmonic frequency is twice of the direct current (DC) power for a narrow pulse train [33]. It indicates that all pulses are well aligned in phase to the 10 GHz signal, so that they all have the maximum contribution to the desired 10 GHz frequency.

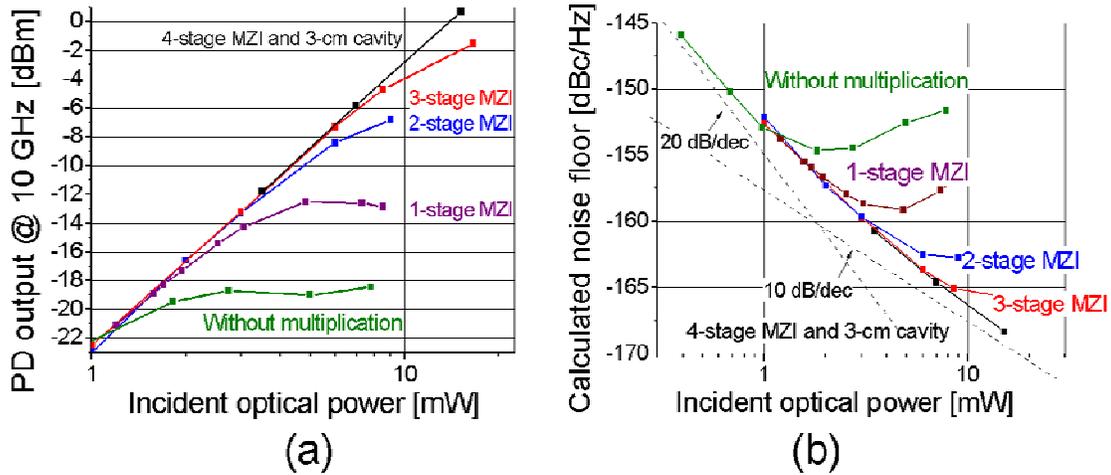

**Figure 6.** (a) Signal size at 10 GHz (i.e., $P_o$) versus incident optical power; (b) predicted single sideband noise floor as given by Eq. 5, where dashed lines are the contributions of the thermal and shot noise with slopes of 20 dB/decade and 10 dB/decade, respectively.

The improvement in 10 GHz power output from the photodiode for different FP and MZI configurations is shown in Figure 6a. Here we plot the measured power in the 10 GHz harmonic from the photodiode versus the incident optical power. As seen, the pulse rate multiplication methods enhance the maximum achieved signal size by nearly 20 dB. With both the 4-stage MZI and 3 cm cavity, we see no evidence of photodiode saturation, and the 10 GHz power increases with the expected slope of 20 dB per decade. In this case, the incident power is limited below 20 mW to avoid damaging the photodiode. With the present photodiodes, there is clearly no advantage to use of a 5-stage MZI or a shorter cavity. The increased signal power should translate into a reduction in the phase noise floor. Figure 6b shows the expected noise floors calculated by use of Eq. 5, with $P_o$ obtained from the data of Figure 6a. The two limiting cases noted previously are seen. At low incident power, the thermal noise dominates and the noise floor improves at a rate of 20 dB per decade. Above a few milliwatts of optical power, the noise floor is limited by shot noise, and the noise floor improvement with incident power is reduced to 10 dB per decade. We note that the estimated noise levels are in agreement with the values given in [33].

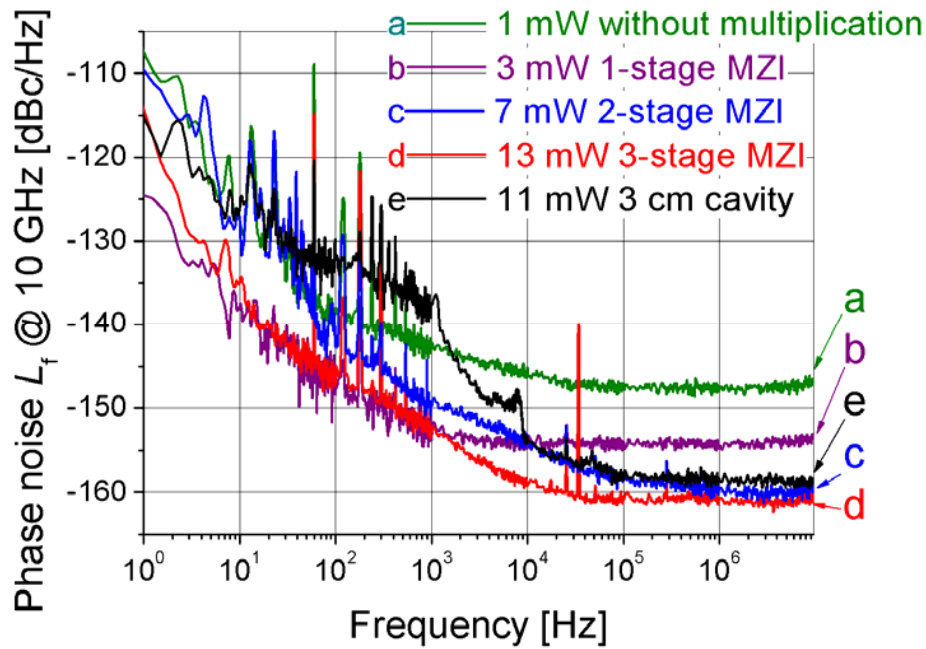

**Figure 7**.  Residual phase noise of MZI and cavity approaches with different incident optical powers.

The results of Fig. 6 indicate that a shot-noise limited phase noise floor of nearly -168 dBc/Hz should be achievable. Using the phase noise measurement system of Fig. 4, we attempt to verify these projections. Figure 7 shows the measured residual single-sideband phase noise of the 10 GHz signal obtained with the previously-discussed setups. The data of Fig. 7 have been corrected by -3 dB to represent the noise of a single system under the assumption of equal noise contributions from the repetition rate multipliers in both arms of the phase measurement system. The lowest phase noise floors are -158 dBc/Hz and -162 dBc/Hz for FP cavity and MZI approaches, respectively, an improvement > 10 dB as compared to no repetition rate multiplication. While not the focus on this work, the additional noise in the frequency range from 1 Hz to 1 kHz is introduced by thermal, mechanical and acoustic fluctuations of the FP filter cavities and the MZI fibers. The variation seen at these frequencies (< 1 kHz) arises from slight differences in the experimental setup and environmental isolation. Improvements can be expected with more attention to these details. In the case of the FP filter cavities (curve e of Fig. 7), the EDFA used before the cavities adds additional noise in the 1 Hz to 10 kHz range. Nonetheless, the close-to-carrier residual phase noise is not seen to be a fundamental limitation.

At Fourier frequencies above 100 kHz, the measured residual phase noise with the MZIs is 3-6 dB higher than the calculated values. This is due primarily to AM-PM conversion, as well as the cavity loss of the 10 GHz bandpass filter (~ 2 dB) and extra noise introduced by the measurement system (including the microwave amplifiers, mixer, baseband amplifier and the Fourier-transform spectrum analyzer). For the FP cavity case (curve e of Fig. 7), the measured phase noise is about 9 dB higher than the calculated value. This is attributed to the amplified spontaneous emission (ASE) of the EDFA required to increase the optical power following the FP cavity filter. We have verified this with an independent measurement of the noise floor of a single PD output with a spectrum analyzer at a frequency of 600 MHz. This noise can be reduced by use of an EDFA with lower noise figure, but it cannot be completely eliminated.

Unfortunately, in all cases the AM-to-PM noise conversion of the photodiodes is not negligible, especially since the commercial mode-locked laser has significant residual intensity noise (RIN ≈ -140 dBc/Hz) at high Fourier frequencies. In order to get the lowest noise floor for the data of Fig. 7, we had to choose a proper working point where $\alpha$ is very low, making the noise converted from laser amplitude noise well below that of other noise sources. One proper working point is about 6 pJ per pulse, corresponding to 1.5 mW without multiplication and 12 mW with a 3-stage MZI. However, for the 4-stage MZI, this working point is beyond the safe operating range of the PD. For the FP cavity filter approach, the proper working point is different because its output is a series of pulses with different energies. While photodiodes with improved AM-to-PM conversion have been recently demonstrated [27], this matter continues to be an important outstanding issue in the generation of the lowest phase noise microwaves.

## 5. Discussion and Conclusions:

The main conclusion of this work is that both approaches to repetition rate multiplication significantly reduce the detection noise floor of a photonic microwave generator, with improvements of nearly 15 dB. However, the MZI approach is clearly better than the cavity approach in regards to power efficiency and the residual noise floor that was realized. In addition, the MZI setup is more robust and is also rather straightforward to realize. For the desired 10 GHz signal, an error of ±2 mm in fiber length corresponds to a phase shift of ±36 degrees or a delay of ±10 ps. Such a level of path length mismatch introduces only about 1 dB of microwave power loss at 10 GHz. We have verified this by using a MZI that incorporates a tunable delay line. Scans of the delay on the order of 1 mm in one arm of a 1-stage MZI introduce no measurable noise level change. In fact, with a 20 GHz sampling oscilloscope monitoring the signal delay variation, it is not difficult to manually splice fiber with a resolution of less than 1 mm. In contrast to what was reported in [22], we find that the desired frequency does not have to be $m2^n$ times of the fundamental repetition frequency, where n and m are integers. A few percent coupling ratio variation can influence the spectrum of the PD output but does not change the residual phase noise level measured with the MZI. It is worth noting that the lowest residual phase noise of the MZI approach is well below that of state-of-the-art photonic microwave generators [8].

The FP cavity-filtering approach does have some potential advantages. First of all, the phase of each pulse is perfectly aligned for all harmonics of the filtered repetition rate. This could be useful for the generation of higher frequencies (e.g. 50 GHz), where the splice length control in the MZI approach will be very difficult. In addition, the mode-filtering cavity can be easily adjusted to achieve any multiplication factor, while the spliced fiber MZI is a fixed system. While noise introduced by the EDFA's results in higher residual phase noise of the FP cavity approach at low Fourier frequencies, its performance can likely be improved with lower noise EDFAs and better environmental noise isolations.

In summary, we have studied two noise floor reduction methods for a 10 GHz fiber-based photonic microwave generator. With the two approaches of MZI and FP mode-filtering cavity, we have decreased the phase noise floor 10 dB and 14 dB, respectively. By comparing the measurement results, we find that the MZI approach is simpler, more robust, and has a lower residual noise for a 10 GHz signal generation. This experiment also proves that the MZI can reduce the detection noise floor without degrading the noise level in low Fourier frequencies. In future work, we will apply the MZI to independent photonic microwave generators as described in Refs [8, 19] to further verify these results and provide 10 GHz signals with absolute phase noise at the lowest levels. While we focus on the application of these techniques to Er:fiber lasers, they should be equally applicable in other laser systems.

## 6. Acknowledgements:

The authors thank A. Hati, C. Nelson, N. Newbury and G. Santarelli for their contributions and comments on this manuscript, and A. Joshi and S. Datta of Discovery Semiconductor for providing the photodiodes. This work was supported by NIST. It is a contribution of an agency of the US government and is not subject to copyright in the USA. Mention of specific products or trade names does not constitute endorsement by NIST.